\documentclass[twocolumn,amsmath,amssymb,prb]{revtex4}
\usepackage{graphicx}
\usepackage{bm}

\begin{document}

\title{Plasmon-mediated superradiance near metal nanostructures}

\author{Vitaliy N. Pustovit$^{1,2}$ and Tigran V. Shahbazyan$^{1}$}

\affiliation{$^{1}$Department of Physics, Jackson State University, Jackson, MS
  39217, USA}

\affiliation{$^{2}$Centre de Recherche Paul Pascal, CNRS, Avenue A. Schweitzer, 33600 Pessac, France} 


\begin{abstract}
We develop a theory of cooperative emission of light by an ensemble of emitters, such as fluorescing molecules or semiconductor quantum dots, located near a metal nanostructure supporting surface plasmon. The primary mechanism of cooperative emission in such systems is resonant energy transfer between emitters and plasmons rather than the Dicke radiative coupling between emitters. We identify two types of plasmonic coupling between the emitters, (i) plasmon-enhanced radiative coupling and (ii) plasmon-assisted nonradiative energy transfer, the competition between them governing the structure of system eigenstates. Specifically, when emitters are removed by more than several nm from the metal surface, the emission is dominated by three superradiant states with the same quantum yield as a single emitter, resulting in a drastic reduction of ensemble radiated energy, while at smaller distances cooperative behavior is destroyed by nonradiative transitions. The crossover between two regimes can be observed in distance dependence of ensemble quantum efficiency. Our numerical calculations incorporating direct and plasmon-assisted interactions between the emitters indicate that they do not destroy the plasmonic Dicke effect.

\end{abstract}


\maketitle

\section{Introduction}
Superradiance of an ensemble of dipoles confined within a limited region in space has been discovered in the pioneering work by Dicke.\cite{dicke-pr54} The underlying physical mechanism can be described as follows. Suppose that a large number, $N$, of dipoles with frequency $\omega_{0}$ are confined in a volume with characteristic size $L$ much smaller than the radiation wavelength $\lambda_{0}=2\pi/\omega_{0}$. Then radiation of an ensemble is a cooperative process in which the emission of a photon is accompanied by virtual photon exchange between individual emitters. This near field radiative coupling between the dipoles leads to  formation of new system eigenstates, each comprised of all individual dipoles. The eigenstates with angular momentum $l=1$ are \textit{superradiant}, i.e.,  their radiative lifetimes are very short, $\sim \tau/N$, where $\tau $ is radiative lifetime of an individual dipole; the remaining states are \textit{subradiant} with much longer decay times,  $\sim \tau (\lambda_{0}/L)^{2}\gg \tau$. 

Since the appearance of Dicke paper, cooperative effects based on Dicke radiative coupling mechanism have been extensively studied in atomic and semiconductor systems (see, e.g., reviews in Refs. \onlinecite{haroche-pr82,andreev-book,brandes-pr05}). Two different decay times corresponding to superradiant and subradiant states were observed in a system of two laser-trapped ions \cite{devoe-prl96} and, more recently, in laterally arranged quantum dots.\cite{scheibner-np07} Other examples of cooperative behavior analogous to the Dicke effect include, e.g., electron tunneling through a system of quantum dots \cite{shahbazyan-prb94,shahbazyan-prb98} and spontaneous phonon emission by coupled quantum dots. \cite{brandes-prl99,brandes-pr05}

\begin{figure}[tb]
  \centering
  \includegraphics[width=0.8\columnwidth]{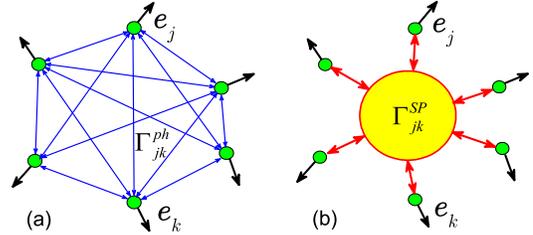}
  \caption{\label{fig:dicke} (Color online) Radiative coupling of emitters in free space (a), and plasmonic coupling of emitters near a metal nanoparticle (b).}
  \end{figure}

Recently, we extended the Dicke effect to plasmonic systems comprised of $N$ dipoles located in the vicinity of a metal nanostructure, e.g., metal nanoparticle (NP), supporting localized surface plasmon (SP).\cite{pustovit-prl09} In such systems, the dominant coupling mechanism between dipoles is \textit{plasmonic} rather  than radiative, i.e., it is based on virtual plasmon exchange (see Fig.\ \ref{fig:dicke}). This plasmonic coupling leads to formation of collective states, similar to Dicke superradiant states, which dominate photon emission. Furthermore, the nanostructure acts as a hub that couples nearby and remote dipoles with about equal strength and hence provides a more efficient hybridization of dipoles compared to radiative coupling. In general, as dipoles orientations in space are non-uniform, there are three superradiant states with total angular momentum $l=1$, each having radiative decay rate $\sim N\Gamma^{r}/3$, where $\Gamma^{r}$ is radiative decay rate of a \textit{single} dipole near a nanostructure (i.e., \textit{with} plasmon enhancement).\cite{pustovit-prl09}

The principal difference between plasmonic and usual (photonic) Dicke effects stems from non-radiative energy transfer between the dipoles and the nanostructure. Let us first outline its role for the case of a \textit{single} dipole near metal NP. When an excited emitter is located close to metal surface,  its energy can be transferred to optically inactive excitations in the metal and eventually dissipated (Ohmic losses). This is described by the \textit{non-radiative decay} rate,  $\Gamma^{nr}\propto d^{-3}$, where $d$ is the dipole--surface separation.\cite{silbey-acp78} Note that very close to metal surface ($\sim 1$ nm), this dependence changes to  $\propto d^{-4}$ due to surface-assisted generation of electron-hole pairs out of the Fermi sea.\cite{persson-prb82,stockman-prb04} As a result, the radiation of a coupled dipole-NP system is governed by a competition between non-radiative losses and plasmon enhancement\cite{moskovits-rmp85}  that determines system quantum efficiency, $Q=\Gamma^{r}/\Gamma$, where $\Gamma = \Gamma^{r}+\Gamma^{nr}$ is the \textit{full} decay rate. Indeed, the radiated energy  is $W=(\hbar kc/2) Q$, $k$ and $c$ being wave vector and speed of light, and its distance dependence follows that of $Q$. Namely, with decreasing $d$, the emission first increases due to plasmon enhancement, and then, at several nm from metal surface, it is quenched due to suppression of $Q$ by non-radiative losses. Both enhancement and quenching were observed in recent experiments on fluorescing molecules attached to a metal NP,\cite{feldmann-prl02,feldmann-nl05,novotny-prl06,sandoghdar-prl06,halas-nl07} and, not too close to NP surface, the distance dependence of \emph{single-molecule} fluorescence \cite{novotny-prl06,sandoghdar-prl06} was found in excellent agreement with single dipole-NP models.\cite{nitzan-jcp81,ruppin-jcp82,ronis-jcp85,chew-jcp87}

When radiation takes place from an \textit{ensemble} of emitters near a metal nanostructure, there are \textit{two} distinct types of plasmon-induced couplings between the emitters. The first is \textit{plasmon-enhanced radiative coupling}, described by radiative decay matrix  $\Gamma_{jk}^{r}$, where indexes $j,k=1,\dots,N$ refer to emitters, that is a straightforward extension of Dicke radiative coupling obtained by incorporating SP local field into the common radiation field. Correspondingly, the eigenstates of $\Gamma_{jk}^{r}$ are superradiant and subradiant states characterized by the strength of their coupling to radiation field. In the ideal case of "point sample," i.e., $kL\ll 1$, the subradiant decay rates are negligibly small and $\Gamma_{jk}^{r}$ essentially has just three non-zero eigenvalues, corresponding to superradiant decay rates, each scaling with $N$ as  $\sim N\Gamma^{r}/3$.\cite{pustovit-prl09}

The second coupling mechanism is \textit{non-radiative energy transfer} between dipoles that takes place in two steps: an excited dipole first transfers its energy to plasmons in nanostructure via its electric field, and then this energy is transferred to another dipole. This process involves plasmons with \textit{all} angular momenta $l$, and it is described by non-radiative decay matrix, $\Gamma_{jk}^{nr}$. Importantly, plasmons with $l>1$ couple to \textit{both} superradiant and subradiant states, so that the eigenstates of \textit{full decay matrix}, $\Gamma_{jk}=\Gamma_{jk}^{r}+\Gamma_{jk}^{nr}$, are not superradiant and subradiant states, but their admixtures. Close to metal surface where non-radiative processes are dominant, $\Gamma_{jk}^{nr}$ prevails over $\Gamma_{jk}^{r}$, and no cooperative behavior is expected. However, when dipoles are removed from the surface by more than several nm, the energy transfer occurs primarily via optically active dipole surface SP and therefore no significant mixing of superradiant and subradiant states takes place and superradiance is intact.

This observation was confirmed by numerical calculation of eigenvalues of $\Gamma_{jk}$, i.e., full decay rates of system eigenstates,  for ensemble of $N$ dipoles randomly distributed in a solid angle around a NP.\cite{pustovit-prl09} Namely, in a wide range of dipole-NP distances, three eigenvalues corresponding to superradiant states are well separated from the rest and scale with $N$ according to $\sim N\Gamma/3$. Since the superradiant states are the optically active ones with radiative decay rate also scaling as $\sim N\Gamma^{r}/3$, their quantum efficiencies essentially \textit{coincide} with $Q$ of single dipole-NP system.  Therefore, in the cooperative regime, the ensemble quantum efficiency, $Q_{ens}$, is thrice that of the single dipole-NP system, 
\begin{equation}
\label{energy}
Q_{ens}\simeq 3Q
\end{equation}
regardless of the ensemble size. Thus, the total radiated energy of the ensemble, $W_{ens}=(\hbar kc/2) Q_{ens}$, is reduced to just $3W$. The remaining energy is trapped by $N-3$ subradiant states and eventually dissipated in the metal rather than being emitted with a much slower rate, as it would be the case in free space. On the other hand, at several nm from metal surface, the non-radiative coupling is dominated by higher $l$ plasmons causing strong mixing of superradiant and subradiant states, i.e., all system eigenstates have comparable quantum efficiencies and  $Q_{ens}\propto N$. Therefore, with increasing distance, $Q_{ens}$, should first exhibit a sharp rise with its slope $\propto N$, and then switch to a more slower $3Q$ dependence.

Indications of such behavior were reported in the recent experiment by  Dulkeith \textit{et al.},\cite{feldmann-nl05} where a systematic study of the ensemble fluorescence vs. distance to metal surface was performed for Cy5 fluorophores attached to Au NP in water. The distance was controlled by varying fluorophores concentration; with increasing concentration,  the linker molecules stretched outwards to accommodate repulsive  dipole-dipole interaction between fluorophores. Therefore, within isolated dipole-NP picture, one would expect that for larger distances, at  which fluorophores concentration was higher than average, the measured quantum efficiency, normalized to some average fluorophores number, should have exceeded the calculated efficiency in single dipole-NP models.\cite{nitzan-jcp81} Instead, with increasing distance, the normalized $Q$, exhibited rapid saturation and, at large distances, was considerably smaller in magnitude than the calculated one.

In the above discussion, we completely ignored interactions between the dipoles. In fact, the role of dipole-dipole interactions in cooperative emission is highly non-trivial since they introduce a disorder into the energy spectrum by causing frequency shifts among randomly distributed in space but otherwise identical emitters.\cite{mukamel-jcp89} Mesoscopic cooperative emission from a disordered system, i.e., for particular disorder realizations rather than its averaged effect, was considered in Ref.~\onlinecite{shahbazyan-prb00} (see also Ref.~\onlinecite{brandes-pr05}). It was found that frequency shifts due to dipole-dipole interactions lift the degeneracy of subradiant states without having significant impact on superradiant states. It was also shown that interactions between collective eigenstates are much weaker than those between individual dipoles, including typical nearest neighbors (i.e., separated by $\sim LN^{-1/3}$), due to  cancellations between dipole-dipole terms among individual pairs with their constituents belonging to different eigenstates.\cite{shahbazyan-prb00} On the other hand, for completely random distribution and high concentration of dipoles, the rare instances of extremely close dipoles (i.e., with separation $\ll LN^{-1/3}$) can prevent the formation of superradiant states.\cite{stockman-prl97}

The role of interactions in plasmon-mediated cooperative emission is characterized by several distinctive features.  First, since the decay matrices $\Gamma_{jk}$ contain plasmon pole, the \textit{relative} strength of dipole-dipole interactions is effectively reduced as compared to purely photonic case.  Second, there are additional corrections to emitters' frequencies, one originating from plasmon-enhanced radiative coupling and another from nonradiative coupling. While neither of those corrections diverges as dipoles approach each other, the latter becomes very large as dipoles approach the metal surface due to contribution of high-$l$ plasmons. Therefore, the actual system eigenstates are determined by \textit{both} the interactions and the energy exchange, and their frequencies and decays rates must be found simultaneously. It is precisely the goal of this paper to calculate the full spectrum of \textit{interacting} emitters near a metal NP.

Specifically, we consider a common situation when emitters, e.g.,  fluorescent molecules or quantum dots, are attached to NP surface via flexible linkers. Typically, fluorophores bound to linker molecules have certain orientation of their dipole moments with respect to NP surface and, due to repulsive interactions, their angular positions are ordered rather then random.\cite{feldmann-nl05}  Therefore, we assume here that angular positions of emitters coincide with the sites of spherical lattice, such as fullerenes. Specifically,  we perform our numerical simulations for C20, C60, and C80 configurations for respective number $N$ of dipoles;  we also study the effect of deviations from ideal lattice. We find that not too close to metal surface, the system eigenstates fall into three groups, each dominated by a particular coupling mechanism: three superradiant states dominated by plasmon-enhanced radiative coupling, one state dominated by direct dipole-dipole interactions, and the rest dominated by non-radiative coupling via NP. Importantly, superradiant states are \textit{not} significantly affected by dipole-dipole interactions whose main effect is a large frequency shift of subradiant state with the smallest decay rate.

We also address the effect of individual emitters' internal nonradiative processes on the ensemble quantum efficiency. Internal relaxation is known to inhibit the photonic Dicke effect if the corresponding decay rate, $\Gamma_{0}^{nr}$, is sufficiently high, i.e., relaxation time is shorter than the lifetime of superradiant state.\cite{andreev-book} However, near metal nanostructure, the plasmon-enhanced radiative decay rate, $\Gamma^{r}$, is significantly larger than $\Gamma_{0}^{nr}$, so that plasmon-mediated superradiance is less sensitive to quantum yield of individual emitters. Specifically, we perform numerical calculations for high-yield and low-yield emitters to show that Eq. (\ref{energy}) holds for both types. More precisely, in cooperative regime, it holds almost exactly for high-yield emitters while for low-yield emitters $Q_{ens}$ is somewhat larger than $3Q$ due to the effective $N$-fold suppression of $\Gamma_{0}^{nr}$ by superradiant states' decay rate.

An obvious application of plasmon-mediated cooperative emission is related to fluorescence of a large but uncertain number of molecules at some average distances from metal nanostructure. For single-molecule case, fluorescence intensity variation with distance was proposed to serve as  nanoscopic ruler,\cite{sandoghdar-nl07} owing to the excellent agreement of measured distance dependences with single-dipole models.\cite{nitzan-jcp81,ruppin-jcp82,ronis-jcp85,chew-jcp87} In the case of molecular layer, the ambiguities caused by uncertain molecules number and their separation from the metal surface prevent, in general, determination of system characteristics from fluorescence variations. However, in cooperative regime,  the ambiguity related to molecules number is removed, and fluorescence intensity is essentially determined by Eq.~(\ref{energy}) with \textit{distance-averaged} single-molecule quantum efficiency.

The paper is organized as follows. In Sec.~\ref{sec:coupling}, the derivation of plasmonic coupling for an ensemble of dipoles distributed near metal NP is given. In Sec.~\ref{sec:energy}, the general expression for radiated energy is obtained. Our numerical results and discussion are presented in Sec.~\ref{sec:disc}. Conclusions and some technical details are provided, respectively, in Sec.~\ref{sec:conc} and in the Appendix.

\section{Plasmonic coupling of radiating dipoles}
\label{sec:coupling}

We consider a system of $N$ emiters, such as fluorescing molecules or quantum dots, with dipole moments $ {\bf d}_{j} = d_{j} {\bf e}_{j}$, where $d_{j}$ and  ${\bf e}_{j}$ are magnitude and orientation, respectively, located at  positions ${\bf r}_{j}$ near a metal NP with radius $R$. Throughout the paper, we  assume that characteristic size of the system (NP+dipoles)  is much smaller than the radiation wavelength, $|{\bf r}_{j}-{\bf r}_{k}|\equiv |r_{jk}|\ll \lambda_{0}$. We also assume that emission events by individual molecules are uncorrelated, i.e., after excitation each molecule relaxes through its own internal nonradiative transitions before emitting a photon. Then the ensemble emission can be described within classical approach by considering dipoles as identical Lorentz oscillators with random initial phases  driven by common electric field, i.e., one created by all dipoles in the presence of metal NP. The frequency-dependent electric field, ${\bf E}({\bf r},\omega )$, satisfies Maxwell's equation
\begin{equation}
\label{maxwell}
\frac{\epsilon ({\bf r},\omega )\omega^{2}}{c^2}  {\bf E}({\bf r},\omega )-{\bm \nabla} \times {\bm \nabla} \times {\bf E}({\bf r},\omega )  = -
\frac{4\pi i \omega}{c^2} {\bf j}({\bf r},\omega ), 
\end{equation}
where dielectric permittivity $\epsilon({\bf r},\omega )$ is that of the metal inside NP, $\epsilon(\omega)$ for $r<R$ and that of outside dielectric $\epsilon_{0}$ for $r>R$. Here ${\bf j}({\bf r},\omega )= -i\int\limits_{0}^{\infty}e^{i\omega t} {\bf j}(t) dt$, is the Laplace transform of dipole current 
\begin{equation}
\label{current}
{\bf j}(t)=q\sum\limits_{j} \dot{d}_{j}(t){\bf e}_{j}\delta ({\bf r}-{\bf r}_{j}),
\end{equation}
where dipole displacements $d_{j}(t)$ are driven by the common electric field at dipoles' positions
\begin{equation}  
\label{displ}      
\ddot {d}_{j} + \omega_{0}^2 d_{j} = \frac{q}{m} {\bf E} ({\bf r}_{j},t) \cdot {\bf e}_{j}
\end{equation}
with the initial conditions (at $t=0$): ${\bf d}_{j}=d_{0} {\bf e}_{j} \sin \varphi_{j}$, 
$\dot {\bf d}_{j}=\omega_{0}d_{0} {\bf e}_{j} \cos \varphi_{j}$, and ${\bf E}=0$  (dot stands for time-derivative). Hereafter, $\omega_{0}$, $q$, $m$, and $\varphi_{j}$ are oscillators' excitation frequency, charge, mass, and initial phase, respectively ($\omega_0=\hbar/md_{0}^{2}$). Closed equations for $d_{j}(\omega)$ can be obtained by using Laplace transform of Eq.~(\ref{displ}) with the above initial conditions  and eliminating ${\bf E}$ from Eqs.~(\ref{maxwell}) and (\ref{displ}). Laplace transform of  Eq.~(\ref{maxwell}) has the form
\begin{align}
\label{maxwell-laplace}
\frac{\epsilon ({\bf r},\omega )\omega^{2}}{c^2}  {\bf E}({\bf r},\omega )-{\bm \nabla} \times {\bm \nabla} \times {\bf E}({\bf r},\omega ) 
\nonumber\\
=\frac{4\pi q}{c^2} \sum_{j} \delta ({\bf r}-{\bf r}_{j}) \bigl [i \omega_{0} d_{0}{\bf e}_{j} \cos\varphi_{j}
\nonumber\\
-\omega^2 d_{j}(\omega ){\bf e}_{j}+ \omega d_{0}{\bf e}_{j} \sin\varphi_{j}\bigr ].
\end{align}
At this point, it is convenient to introduce normalized displacements
\begin{align}
 v_{j}(\omega)=d_{j}(\omega)/d_{0} - i\left (\omega_{0}/\omega^2\right ) \cos \varphi_{j} - \omega^{-1}\sin \varphi_{j}
 \end{align}
and the solution of Eq.~(\ref{maxwell-laplace})  reads
\begin{eqnarray}
\label{electric}
{\bf E}({\bf r},\omega ) = \frac{4\pi d_{0}q \omega^{2}}{c^2} \sum_{j}{\bf G}({\bf r},{\bf r}_{j},\omega) \cdot {\bf e}_{j} v_{j}, 
\end{eqnarray}
where ${\bf G}({\bf r},{\bf r}',\omega)$ is the electric field Green diadic \textit{in the presence of} NP. From Eq.~(\ref{displ}), for photon frequency close to dipoles frequency, $\omega \approx \omega_{0}$, we obtain a coupled system of equations for normalized displacements,
\begin{eqnarray}
\label{system}
\sum_{k} \Bigl[(\omega_{0}-\omega ) \delta_{jk} +  \Sigma _{jk}\Bigr] v_{k} = 
\frac{-i}{2} e^{-i\varphi_{j}},
\end{eqnarray}
where $\Sigma _{jk}=\Delta_{jk}-\frac{i}{2}\Gamma_{jk}$ is the complex \emph{self-energy matrix}, given by
\begin{eqnarray}
\label{self-general}
\Sigma _{jk}(\omega)= - \frac{2\pi q^2 \omega_{0}}{m c^2} \, {\bf e}_{j}\cdot  {\bf G}({\bf r}_{j},{\bf r}_{k};\omega) \cdot {\bf e}_{k}.
\end{eqnarray}
The system Eq.~(\ref{system}) determines eigenstates of the ensemble of $N$ dipoles in common radiation and plasmon field, while complex eigenvalues of the self-energy matrix give eigenstates frequency shifts with respect to $\omega_{0}$ and their decay rates. In the absence of NP, real and imaginary parts of the self-energy matrix are dipole-dipole interaction and radiation coupling between dipoles $j$ and $k$, given by (in lowest order in $kr_{jk}$)
\begin{align}
\label{dipole-dipole}
&\Delta_{jk}^{0}=\frac{3\Gamma_{0}^{r}}{4(kr_{jk})^3} 
\biggl[({\bf e}_{j} \cdot {\bf e}_{k})-\frac{3({\bf e}_{j}\cdot {\bf r}_{jk})
({\bf e}_{k} \cdot {\bf r}_{jk})}{r_{jk}^2}\biggr],
\nonumber\\
&\Gamma_{jk}^{0}=\Gamma_{0}^{r}\, {\bf e}_{j} \cdot {\bf e}_{k},
\end{align}
where 
\begin{equation}
\label{decay_rad0}
\Gamma_{0}^{r}=\frac{2k q^2\omega_{0}}{3 m c^2}=\dfrac{2 \mu^{2} k^{3}}{3 \hbar\epsilon_{0}},
\end{equation}
is the radiative decay rate of a dipole in a dielectric medium, $\mu=qd_{0}$ is the dipole moment, and $k=\sqrt{\epsilon_{0}}\omega/c$ is the wave vector. The eigenstates of \textit{photonic} decay matrix $\Gamma_{jk}^{0}$ are superradiant and subradiant states, the former having  decay rate $\sim N\Gamma_{0}^{r}$. In the case when all dipoles are aligned, there is only one superradiant state that couples to radiation field, while for general dipole orientations there are three such states with angular momentum $l=1$. Note that in the longwave approximation used here, the decay rates of subradiant states vanish ("point sample").

In the presence of metal nanostructure, the system eigenstates are determined by the full Green diadic in self-energy matrix Eq.~(\ref{self-general}). In the case of spherical NP, the longwave approximation for  ${\bf G}(r_{j},r_{k};\omega)$ can be easily found.\cite{ruppin-jcp82,chew-jcp87,stockman-njp08} The details are given in the Appendix, and the result reads
\begin{align}
\label{self}
\Sigma_{jk}(\omega)=&\Delta_{jk}^{0}-\frac{3\Gamma_{0}^{r}}{4k^3} \sum_{l} \alpha_{l} T_{jk}^{(l)} 
\\
-&\frac{ i}{2}\Gamma_{0}^{r}\Bigl [({\bf e}_{j} \cdot {\bf e}_{k})-\alpha_{1} \bigl[K_{jk}^{(1)} + h.c.\bigr] +|\alpha_{1}|^2 T_{jk}^{(1)}\Bigr ],
\nonumber
\end{align}
where 
\begin{equation}
\label{alpha_l}
\alpha_{l}(\omega)=\frac{R^{2l+1}\left [\epsilon(\omega) -\epsilon_0\right ]}{\epsilon(\omega) +(1+1/l)\epsilon_0}
\end{equation}
 is  NP $l$-pole polarizability. The matrices $K_{jk}^{(l)}$ and $T_{jk}^{(l)}$ are defined as
\begin{align}
K_{jk}^{(l)}=\frac{4\pi}{2l+1} \sum_{m=-l}^{l} [{\bf e}_{j} \cdot {\bm \psi}_{lm}({\bf r}_{j})] [{\bf e}_{k} \cdot {\bm \chi}_{lm}^{*}({\bf r}_{k})],
\label{K}
\\
T_{jk}^{(l)}=\frac{4\pi}{2l+1} \sum_{m=-l}^{l} [{\bf e}_{j} \cdot {\bm \psi}_{lm}({\bf r}_{j})] [{\bf e}_{k} \cdot {\bm \psi}_{lm}^{*}({\bf r}_{k})],
\label{T}
\end{align}
where 
\begin{equation}
{\bm \psi}_{lm}({\bf r})={\bm \nabla} \left [\dfrac{Y_{lm}(\hat{\bf r})}{r^{l+1}}\right ],
~~
{\bm \chi}_{lm}({\bf r})={\bm \nabla} \left [r^{l} Y_{lm}(\hat{\bf r})\right ],
\end{equation}
and $Y_{lm}(\hat{\bf r})$ are spherical harmonics. For $l=1$, these matrices can be evaluated as
\begin{align}
\label{K1}
K^{(1)}_{jk}=&\frac{1}{r^3_j} \Bigl[({\bf e}_{j}\cdot {\bf e}_{k}) -3({\bf e}_{j}\cdot \hat{\bf r}_{j}) (\hat{\bf r}_{j}\cdot {\bf e}_{k})\Bigr], 
\\
\label{T1}
T^{(1)}_{jk}=&\frac{1}{r^3_j r^3_k} \Bigl[({\bf e}_{j}\cdot {\bf e}_{k})
- 3({\bf e}_{j}\cdot \hat{\bf r}_{j}) ({\bf e}_{k}\cdot \hat{\bf r}_{j})
\\
-&3({\bf e}_{j}\cdot \hat{\bf r}_{k}) ({\bf e}_{k}\cdot \hat{\bf r}_{k})
+9({\bf e}_{k}\cdot \hat{\bf r}_{k})({\bf e}_{j}\cdot \hat{\bf r}_{j})(\hat{\bf r}_{j}\cdot \hat{\bf r}_{k})\Bigr], 
\nonumber
\end{align}
where  we have used identities
\begin{align}
\label{ident}
&\sum_{m=-1}^{1} {\bm \psi}_{1m} ({\bf r}_k) Y^{*}_{1m}(\hat{\bf r}) = \frac {3}{4 \pi r^{3}_k} \bigl[\hat{\bf  r} - 3 \hat{\bf r}_k 
(\hat{\bf r}\cdot \hat{\bf r}_k) \bigr],
\nonumber\\
&\sum_{m=-1}^{1} {\bm \chi}_{1m} ({\bf r}_k) Y_{1m}^{*}(\hat{\bf r}) = \frac {3} {4 \pi} \hat{\bf r}.
\end{align}
For dipoles oriented normally with respect to NP surface (${\bf e}_j={\bf \hat r}_j$), we obtain
\begin{align}
\label{KT-perp}
K_{jk}^{(1)}=-\frac{2}{r_j^3} ({\bf e}_{j} \cdot {\bf e}_{k}),
~
T^{(1)}_{jk}=\frac{4}{r_j^3 r_k^3} ({\bf e}_{j} \cdot {\bf e}_{k}),
\end{align}
and for parallel orientation (${\bf e}_j\cdot{\bf \hat r}_j=0$), we similarly get
\begin{align}
\label{KT-par}
K_{jk}^{(1)}=\frac{1}{r_j^3} ({\bf e}_{j} \cdot {\bf e}_{k}),
~~
T^{(1)}_{jk}=\frac{1}{r_j^3 r_k^3} ({\bf e}_{j} \cdot {\bf e}_{k}).
\end{align}
The \emph{decay matrix}, $\Gamma_{jk}=-2{\rm Im} \Sigma_{jk}$, can be decomposed into radiative and nonradiative terms, $\Gamma_{jk}=\Gamma_{jk}^{r}+\Gamma_{jk}^{nr}$, as follows
\begin{align}
\label{width}
&\Gamma_{jk}^{r}= \Gamma_{0}^{r} \Bigl[({\bf e}_{j} \cdot {\bf e}_{k})-\alpha'_{1}\bigl [K_{jk}^{(1)} + h.c.\bigr ] 
+ |\alpha_{1}|^2 T_{jk}^{(1)}\Bigr],
\nonumber\\
&\Gamma_{jk}^{nr}=\frac{3\Gamma_{0}^{r}}{2k^{3}} \sum_l \alpha''_l T_{jk}^{(l)}.
\end{align}
The diagonal elements $\Gamma_{jj}^{r}$ and $\Gamma_{jj}^{nr}$ describe, respectively, plasmon-enhanced radiative decay rate of an \textit{isolated} dipole near a NP, and nonradiative transfer of its energy to electronic excitations in metal. The non-diagonal elements of $\Gamma_{jk}^{r}$ describe \textit{plasmon-enhanced radiative coupling} that generalizes the Dicke mechanism responsible for cooperative emission by incorporating local field enhancement into near field radiative coupling. On the other hand, the non-diagonal terms in $\Gamma_{jk}^{nr}$ describe \textit{nonradiative energy transfer} between dipoles mediated by NP plasmons. The latter coupling is absent in the photonic Dicke effect but it plays important role in the plasmonic Dicke effect, as we will see below.

Since the numerical calculations below are carried for normal dipoles orientations, we provide here the corresponding expressions for self-energy matrix $\Sigma_{jk}$. The decay matrix has the form 
\begin{align}
\label{gamma-perp}
&\Gamma_{jk}^{r}= \Gamma_{0}^{r} \left [1 + 2 \alpha'_{1} \left (\frac{1}{r_j^{3}}+\frac{1}{r_k^{3}}\right ) + \frac{4 | \alpha_{1}|^2}{r_j^{3}r_k^{3}}\right ]\cos \gamma_{jk} ,
\nonumber\\
&\Gamma_{jk}^{nr}= \frac{3\Gamma_{0}^{r}}{2k^{3}} \sum_l \frac{ \alpha''_l (l+1)^2 }{r_{j}^{l+2} r_{k}^{l+2}}P_l(\cos \gamma_{jk}),  
\end{align}
where $P_{l}$ is Legendre polynomial and $\gamma_{jk}$ is the angle between positions of dipoles $j$  and $k$ measured from  NP center (here $\cos\gamma_{jk}={\bf e}_{j}\cdot {\bf e}_{k}$). The real part of self-energy matrix has the form
\begin{align}
\label{delta-perp}
\Delta_{jk}= \Delta_{jk}^{0} + & \Gamma_{0}^{r} \biggl [\alpha''_1 \left (\frac{1}{r_{j}^3} + \frac{1}{r_{k}^3} \right ) \cos \gamma_{jk} 
\nonumber\\
& - \frac{3}{4k^3} 
\sum_l \frac{ \alpha'_l (l+1)^2 } {r_{j}^{l+2} r_{k}^{l+2} }P_l (\cos \gamma_{jk}) \biggr ],
\end{align}
where the second term describes NP-induced interactions. The latter in turn consists of two terms, first coming from plasmon-enhanced radiative coupling and second coming from non-radiative coupling.

Note that both NP-induced terms are weaker than their counterparts in $\Gamma_{jk}$ while having same symmetry and therefore is not expected to significantly alter the eigenstates. On the other hand, the dipole-dipole interaction term, Eq.~(\ref{dipole-dipole}),  has different symmetry and can become large for two dipoles in a close proximity to each other. The effect of interactions on cooperative emission is studied in Sec. IV.

%
\section{Radiated energy}
\label{sec:energy}

In this section, we derive general expression for total energy radiated by an ensemble of dipoles near a metal nanostructure. The radiated energy in the unit frequency interval is obtained by integrating the far field ($r\rightarrow \infty$) spectral intensity over solid  angle\cite{novotny-book}
\begin{eqnarray}
\label{energy-general}
\frac{dW(\omega)}{d\omega}= \frac{c\epsilon_{0}}{4\pi^{2}}\int \left |{\bf E}({\bf r},\omega)\right |^{2}r^{2}d\Omega,
\end{eqnarray}
and then averaging the result over initial random phases of individual dipoles, $\varphi_{j}$. The electric field ${\bf E}({\bf r},\omega)$ is given by Eq.~(\ref{electric}), where $v_{j}$ is the solution of Eq.~(\ref{displ}).  Then the energy density takes the form
\begin{align}
\label{energy-flow}
\frac{dW}{d\omega}=&\frac {4 r^2 \epsilon_0 \mu^{2} \omega_0^4}{c^3} 
\\
&\times \sum_{jk} \int d\Omega v_j v^{*}_k \bigl[{\bf G}({\bf r},{\bf r}_j) \cdot {\bf e}_j\bigr] \cdot \bigl[{\bf G}^{*}({\bf r},{\bf r}_k) \cdot {\bf e}_k\bigl],
\nonumber
\end{align}
with ${\bf G}({\bf r},{\bf r}_{j},\omega)={\bf G}^{0}({\bf r},{\bf r}_{j},\omega)+{\bf G}^{s}({\bf r},{\bf r}_{j},\omega)$ replaced by its far field asymptotics  (see Appendix),
\begin{eqnarray}
\label{Green-farfield}
G_{\mu\nu}(r,r_j)= \frac{e^{ik r}}{4\pi r} 
\Bigl[ \delta_{\mu \nu } 
-\frac{4 \pi}{3} \sum_{m}  \hat{\bf r}_{\mu} Y_{1m}(\hat{\bf r})\chi_{1m}^{\nu *}({\bf r}_j) 
\nonumber\\
-\frac{4 \pi}{3}\tilde \alpha_1 r \sum_{m} \bigl[\nabla_{\mu} Y_{1m}(\hat{\bf r}) \bigr] \psi_{1m}^{\nu *}({\bf r}_{j}) \Bigr],
\end{eqnarray} 
where first two terms come from the free space part,  ${\bf G}^{0}$, and the last term comes from the scattered part, ${\bf G}^{s}$, of the Green diadic. The angular integral in Eq.~(\ref{energy-general}) can be performed using the relations  Eq.~(\ref{ident}). The free space contribution yields 
\begin{align}
4 \pi r^2  \int  d\Omega \bigl[{\bf G}^{0}({\bf r},{\bf r}_j) \cdot {\bf e}_j\bigr] \cdot \bigl[{\bf G}^{0*}({\bf r},{\bf r}_k)\cdot {\bf e}_k\bigr]=
\frac {2}{3}  ({\bf e}_j \cdot {\bf e}_k).
\end{align} 
The other integrals in the product Eq.~(\ref{energy-flow})  are evaluated using the relations
\begin{align}
\left (\frac{4 \pi r}{3}\right )^2 \sum_{mm'}
& 
\bigl[{\bm \psi^{*}_{1m}} ({\bf r}_j) \cdot {\bf e}_j\bigr] \bigl[{\bm \psi_{1m'}} ({\bf r}_k) \cdot {\bf e}_k\bigr]
\\
&
\times \int \frac{d\Omega}{4\pi} {\bm \nabla} Y_{1m} (\hat{\bf r}) \cdot {\bm \nabla} Y^{*}_{1m'} (\hat{\bf r}) = \frac {2}{3} T^{(1)}_{jk},
\nonumber
\end{align}
and
\begin{align}
\frac{4 \pi r}{3} \sum_m \int \frac{d\Omega}{4\pi} \bigl[{\bf e}_k\cdot {\bm \nabla} Y_{1m}(\hat{\bf r})\bigr]
\bigl[{\bm \psi}_{1m}^{*} ({\bf r}_j) \cdot {\bf e}_j\bigr]
= \frac {2}{3}  K^{(1)}_{jk},
\end{align}
yielding
\begin{align}
 &\int \frac{d\Omega}{4\pi}  \bigl[{\bf G}({\bf r},{\bf r}_j) \cdot {\bf e}_j\bigr] \cdot \bigl[{\bf G}^{*}({\bf r},{\bf r}_k) \cdot {\bf e}_k\bigr]
\\  
&=\frac{1}{(4\pi r)^2} \frac {2}{3} \Bigl[({\bf e}_j \cdot {\bf e}_k) - 
\alpha_{1} K_{jk}^{(1)} - \alpha^{*}_{1} K_{kj}^{(1)}+
|\alpha_{1}|^2 T_{jk}^{(1)} \Bigr].
\nonumber
\end{align}
The energy density then takes the form
\begin{eqnarray}
\label{flow-simplify}
\frac{dW(\omega )}{d\omega}= \frac{\sqrt{\epsilon_0}\hbar \omega_0}{\pi} \sum_{jk} v_j v^{*}_k A_{jk},
\end{eqnarray}
where $A_{jk}=\Gamma_{0}^{r} \Bigl[({\bf e}_j \cdot {\bf e}_k)  
-\tilde \alpha_{1} K_{jk}^{(1)} - \tilde \alpha^{*}_{1} K_{kj}^{(1)} + |\tilde \alpha_{1}|^2 T_{jk}^{(1)} \Bigr]$. Matrix $A_{jk}$ is not symmetrical, however only its symmetrical part, equal to $\Gamma^{r}_{jk}$, contributes to the final expression. The solution of Eq.~(\ref{system}) can be presented as
\begin{eqnarray}
\label{v_j}
v_j=-\frac{i}{2}\sum_k \biggl[ \frac{1}{\omega_{0}-\omega  +  \hat{\Sigma} }\biggr]_{jk}  e^{-i \varphi _k},
\end{eqnarray}
and after averaging out over the initial random phases $\varphi_j$, we finally obtain
\begin{align}
\label{final-flow}
\frac{dW(\omega )}{d\omega}= \frac{\hbar kc}{4\pi}\, {\rm Tr} \left[ \frac{1}{\omega-\omega_{0}  -  \hat{\Sigma}^{\dagger} } \hat{\Gamma}^{r} \frac{1}{\omega-\omega_{0}  -  \hat{\Sigma}}\right],
\end{align}
where  trace is taken over the indexes $(jk)$. The analysis of this expression and the results of numerical calculations will be presented in the next section.


\section{Discussion and numerical results}
\label{sec:disc}

Let us start with a \textit{single} dipole located at $r_{0}$ near a metal NP. In this case, the self-energy is a complex number, $\Sigma=\Delta -\frac{i}{2}\Gamma$, where $\Delta=\Delta_{jj}$ and  $\Gamma=\Gamma^{r}+\Gamma^{nr}=\Gamma_{jj}$ are  single-dipole energy shift and decay rate, respectively. For normal ($s=\perp$) and parallel ($s=\parallel$) dipole orientations, using Eqs.~(\ref{KT-perp}) and (\ref{KT-par}), these are given by\cite{nitzan-jcp81}
\begin{align}
\label{classic}
&\Gamma^{r}=\Gamma_{0}^{r}\left |1+\frac{a_{s}\alpha_{1}}{r_{0}^{3}}\right |^{2},
~
\Gamma^{nr}= \frac{3\Gamma_{0}^{r}}{2k^{3}} \sum\limits_l \frac{b_{s}^{(l)} \alpha''_l }{r_{0}^{2l+4}},
\nonumber\\
&\Delta=\Gamma_{0}^{r}\left (\frac{a_{s}\alpha''_{1}}{r_{0}^{3}}
-\frac{3}{4k^{3}} \sum\limits_l \frac{b_{s}^{(l)} \alpha'_l }{r_{0}^{2l+4}}\right ),
\end{align}
 with $a_{\perp}=2$, $b_{\perp}^{(l)}=(l+1)^2$ and $a_{\parallel}=-1$,  $b_{\parallel}^{(l)}=l(l+1)/2$.  Note that both  terms in $\Delta$ are smaller than their counterparts $\Gamma^{r}$ and $\Gamma^{nr}$ due to plasmon pole in the imaginary part of NP polarizability $\alpha''_{l}$. Radiated energy of single dipole-NP system, obtained by frequency integration of Eq.~(\ref{final-flow}), is given by
\begin{eqnarray}
\label{total_energy-single}
W=\frac{\hbar kc}{2}\,\frac{\Gamma^{r}}{\Gamma+\Gamma_{0}^{nr}}=\frac{\hbar kc}{2}\,Q,
\end{eqnarray}
 where we included internal molecular relaxation rate, $\Gamma_{0}^{nr}$, into quantum  efficiency $Q$. For $N$ \textit{uncoupled} dipoles, i.e., for purely diagonal $\Sigma_{jk}=\delta_{jk}\left (\Delta- \frac{i}{2}\Gamma\right )$,  Eq.~(\ref{final-flow}) decouples into sum of $N$ independent terms, yielding $W_{ens}=NW$. 
 
In the presence of inter-dipole coupling, the system eigenstates, $|J\rangle$, are those of the self-energy matrix, Eq.~(\ref{self}). The corresponding eigenvalues are complex,  $\hat{\Sigma}|J\rangle =(\Delta_{J}-\frac{i}{2}\Gamma_{J})|J\rangle$, where $\Delta_{J}$ is frequency shift of collective eigenstate $|J\rangle$ relative to $\omega_{0}$ and $\Gamma_{J}$ is its decay rate. The molecular relaxation  can be accounted for by adding  to $\Sigma_{jk}$ a diagonal term, $-\frac{i}{2}\delta_{jk}\Gamma_{0}^{nr}$. Then, after frequency integration of Eq.~(\ref{final-flow}), the ensemble radiated energy takes the form 
\begin{equation}
\label{Q-ens}
W_{ens}=\frac{\hbar kc}{2}\, Q_{ens},
~~~Q_{ens}=\sum\limits_{J}\dfrac{\Gamma_{J}^{r}}{\Gamma_{J}+\Gamma_{0}^{nr}},
\end{equation}
where $\Gamma_{J}^{r}=\langle J|\hat{\Gamma}^{r}|J\rangle$ is radiative decay rate of state $|J\rangle$.

In the photonic Dicke effect, superradiant and subradiant states are eigenstates of the radiative decay matrix $\Gamma_{jk}^{0}$ obtained from the free space Green diadic. Similarly, in the plasmonic Dicke effect, superradiant states are eigenstates of \textit{plasmon-enhanced} radiative decay matrix $\Gamma_{jk}^{r}$. Let us illustrate the emergence of plasmon-mediated superradiance for a simple case when all dipoles are at the same distance from NP surface and are oriented normal or parallel to it. Then it is easy to see that the corresponding decay matrix, $\Gamma_{jk}^{r}=\Gamma^{r}{\bf e}_{j}\cdot {\bf e}_{k}$ with $\Gamma^{r}$ given by Eq.~(\ref{classic}), has just three nonzero eigenvalues. Indeed, let us introduce new decay matrices as
 \begin{equation}
\gamma_{\mu\nu}^{r}=\dfrac{N\Gamma^{r}}{3} B_{\mu\nu},
\end{equation}
where 
\begin{equation}
B_{\mu\nu}=\dfrac{3}{N}\sum_{j} e_{j}^{\mu}e_{j}^{\nu} 
\end{equation}
is $3\times 3$  matrix in coordinate space with ${\rm Tr}\hat{B}=3$. It is easy to see that ${\rm Tr}\bigl [ (\hat{\Gamma}^{r} )^{n}\bigr]={\rm Tr}\bigl [ (\hat{\gamma}^{r} )^{n}\bigr]$ for any integer $n$, i.e., the $N\times N$ matrix $\Gamma_{jk}^{r}$ has only three non-zero eigenvalues coinciding with those of matrix $\gamma_{\mu\nu}^{r}$
\begin{equation}
\label{gamma-coop}
\Gamma_{\mu}^{r}=\dfrac{N\Gamma^{r}}{3}\,\lambda_{\mu},
\end{equation}
where $\lambda_{\mu}\sim 1$ are eigenvalues of $B_{\mu\nu}$. Note that the decay rates of the remaining $N-3$ subradiant  states vanish in the long wave approximation used here; they acquire finite values in the next order in $(kr_{0})^{2}$.

Let us turn to non-radiative coupling, described by matrix $\Gamma_{jk}^{nr}$. Its diagonal elements, $\Gamma_{jj}^{nr}$, describe nonradiative energy exchange between excited dipole and NP plasmon modes with all angular momenta, as indicated by polarizabilities $\alpha''_{l}$ in Eq.~(\ref{gamma-perp}). The non-diagonal elements of $\Gamma_{jk}^{nr}$ describe a process by which a plasmon nonradiatively excited in the NP by dipole $k$ transfers its energy to another dipole $j$. In general, due to high-$l$ plasmons involved in nonradiative coupling, the eigenstates of $\Gamma_{jk}^{nr}$ are \textit{different} from  those of plasmon-enhanced radiative coupling $\Gamma_{jk}^{r}$ which contains only dipole ($l=1$) plasmon mode. Therefore, the eigenstates of full decay matrix $\Gamma_{jk}=\Gamma_{jk}^{nr}+\Gamma_{jk}^{nr}$ are \textit{not} pure superradiant and subradiant states but their admixtures. However, the high-$l$ plasmon contribution to $\Gamma_{jk}^{nr}$ is significant only at very small $d$ [see, e.g., Eq.~(\ref{gamma-perp})] while for $d$ larger than several nm $\Gamma_{jk}^{nr}$ is dominated by the $l=1$ term.  In fact, in a wide range $d$, nonradiative coupling between dipoles is mainly through the optically active \textit{dipole} plasmon mode that does not cause mixing between superradiant and subradiant states. Namely, it can be easily seen from Eq.~(\ref{gamma-perp}) that the eigenstates of the $l=1$ term in $\Gamma_{jk}^{nr}$ are the same  as those of $\Gamma_{jk}^{r}$ so the corresponding eigenvalues are similarly given by $\Gamma_{\mu}^{nr}=\left (N\Gamma^{nr}/3\right )\lambda_{\mu}$. Thus, the superradiant quantum efficiency
\begin{equation}
Q_{\mu}=\frac{\Gamma_{\mu}^{r}}{\Gamma_{\mu}+\Gamma_{0}^{nr}}=\dfrac{\Gamma^{r}}{\Gamma+3\Gamma_{0}^{nr}/N\lambda_{\mu}}
\end{equation}  
only weakly depends on $N$. Therefore, the sum in Eq.~(\ref{Q-ens}) includes just three terms, yielding
\begin{equation}
\label{total_energy-ens}
W_{ens}=\dfrac{\hbar kc}{2} \, Q_{ens}=
\dfrac{\hbar kc}{2}\sum\limits_{\mu=1}^{3}\dfrac{\Gamma^{r}}{\Gamma+3\Gamma_{0}^{nr}/N\lambda_{\mu}}.
\end{equation}
For high-yield (small $\Gamma_{0}^{nr}$) emitters, we obtain Eq.~(\ref{energy}) and hence $W_{ens}\approx 3W$. In contrast, the radiated \textit{power},
\begin{equation}
\label{power-ens}
P_{ens}=\dfrac{\hbar kc}{2}\sum\limits_{\mu=1}^{3}Q_{\mu}\Gamma_{\mu}^{r} \simeq N \left (\dfrac{\hbar kc}{2}\,Q\Gamma^{r}\right )=NP,
\end{equation}
scales with the ensemble size due to shorter (by factor $N/3$) radiative lifetime of superradiant states. 

For low-yield emitters (large $\Gamma_{0}^{nr}$), the relation Eq.~(\ref{energy}) holds only approximately. However, it is evident from comparison of  Eqs.~(\ref{total_energy-single}) and (\ref{total_energy-ens}) that here the relative effect of internal relaxation is much weaker than for usual cooperative emission.  Numerical results for both high-yield and low-yield emitters are presented below.

Let us now turn to the role of interactions between dipoles in the ensemble, which is the main subject of this paper. Interactions play critical role in cooperative emission since they introduce a disorder into system energy spectrum by causing random shifts of individual dipole frequencies.\cite{mukamel-jcp89,shahbazyan-prb00,stockman-prl97} In the conventional cooperative emission, the main disorder effect is to split the narrow subradiant peak in the ensemble emission spectra.\cite{shahbazyan-prb00} In the presence of metal nanostructure,  radiation of subradiant states is expected to be quenched by much faster nonradiative losses in the metal. The crucial question is, however, whether interactions between closely spaced individual dipoles can significantly alter the structure of collective eigenstates. In the remaining part  of the paper, we present the results of our numerical simulations of cooperative emission fully incorporating both direct  and plasmon-mediated interactions.

We consider  an ensemble of $N$ molecular dyes attached to an Ag spherical particle with radius $R=20$ nm via molecular linkers with approximately same length. The system is embedded in aqueous solution with dielectric constant $\epsilon_0=1.77$, and two types of dyes with quantum efficiencies $q=0.3$ and $q=0.95$ are used in the calculations.  A distinguishing feature of this system is a strong effect of interactions on its geometry.\cite{feldmann-nl05} The flexible linker molecules hold the attached dyes with certain orientation of their dipole moments, so that repulsive inter-molecule interactions compel the dyes to form a spatially ordered structure on spherical surface. In our simulations, the dyes with normal dipole orientations were located at the sites of spherical lattice, specifically, fullerenes C20, C32, C60, and C80, and, in some calculations,  we included random deviations from the ideal lattice positions. The system eigenstates are found by numerical diagonalization of self-energy matrix, $\Sigma_{jk}=\Delta_{jk}-\frac{i}{2}\Gamma_{jk}$, with its real and imaginary parts given by Eqs.~(\ref{delta-perp}) and (\ref{gamma-perp}), respectively. Calculations were carried at the SP energy of 3.0 eV, the size-dependent Landau damping was incorporated for all plasmon modes, and NP polarizabilities, Eq.~(\ref{alpha_l}),  with angular momenta up to $l=30$, were calculated using the experimental bulk Ag complex dielectric function. 

  \begin{figure}
  \centering
  \includegraphics[width=0.8\columnwidth]{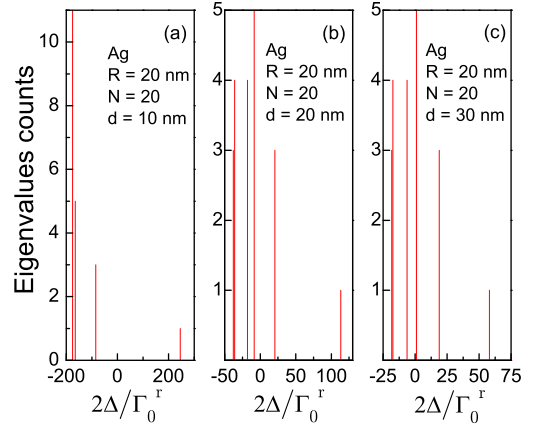}
  \caption{\label{fig:hist_delta}(Color online) Distribution of energy shifts for 20 dipoles in C20 configuration around Ag NP at several distances to its surface.}
  \end{figure}
  \begin{figure}
  \centering
  \includegraphics[width=0.8\columnwidth]{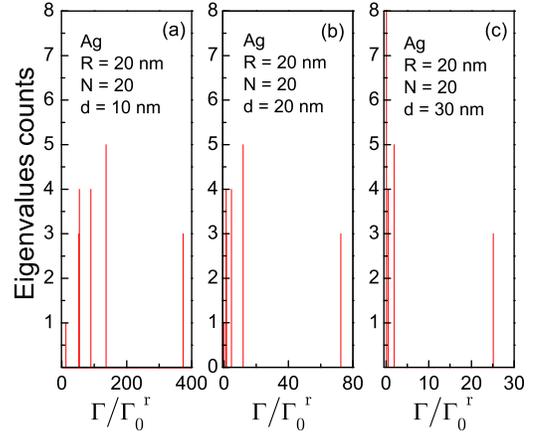}
  \caption{\label{fig:hist_gamma} (Color online) Distribution of decay rates for 20 dipoles in C20 configuration around Ag NP at several distances to its surface.}
  \end{figure}

Figures \ref{fig:hist_delta} and \ref{fig:hist_gamma} show distribution of real and imaginary parts of complex eigenvalues of $\Sigma_{jk}$ for $N=20$ molecules at the sites of C20 fullerene at three different molecule-surface distances. The system spectrum represents several sets of degenerate eigenvalues indicating a high degree of lattice symmetry. For all distances, the histograms show a \textit{single} eigenvalue with a large positive energy shift (Fig.~\ref{fig:hist_delta}), which corresponds to the direct dipole-dipole interaction between nearest-neighbor molecules. On the other hand, there are \textit{three} degenerate eigenvalues with the largest decay rate, corresponding to predominantly \textit{superradiant} states while the smaller decay rates are those of predominantly subradiant states (Fig.~\ref{fig:hist_gamma}). With increasing distance, the mixing between superradiant and subradiant states decreases  and, for $d=30$ nm, decay rates of all but three eigenstates nearly vanish; note that in our approximation, pure subradiant states should have zero decay rate. Such behavior is due to diminishing contribution of higher-$l$ plasmons at larger distances [see Eqs.~((\ref{gamma-perp}) and \ref{delta-perp})]. Importantly, direct interactions between close molecules result only in energy shift of subradiant states, without affecting  superradiant states. We therefore conclude that dipole-dipole interactions do \textit{not} destroy cooperative emission in plasmonic systems.

  \begin{figure}
  \centering
  \includegraphics[width=0.8\columnwidth]{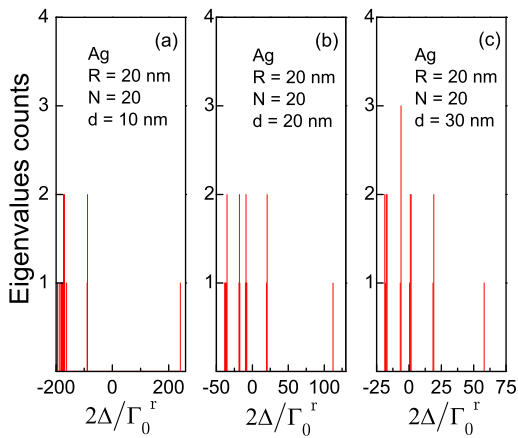}
  \caption{\label{fig:hist_delta-disorder}(Color online) Distribution of energy shifts for 20 dipoles in C20 configuration around Ag NP at several average (with  10\% fluctuations) distances to its surface.}
\end{figure}
  \begin{figure}
  \centering
  \includegraphics[width=0.8\columnwidth]{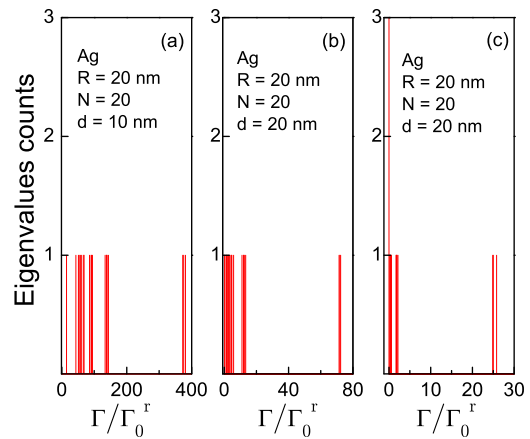}
  \caption{\label{fig:hist_gamma-disorder} (Color online) Distribution of decay rates for 20 dipoles in C20 configuration around Ag NP at several average (with  10\% fluctuations) distances to its surface.}
\end{figure}

This main conclusion remains unchanged when fluctuations (up to 10\%) of molecules positions in radial direction are included into simulations (Figs.~\ref{fig:hist_delta-disorder} and \ref{fig:hist_gamma-disorder}). The spatial disorder lifts lattice symmetry, so that superradiant states now have different, however close, decay rates. Note that without interactions, i.e.,  when molecules angular positions are completely random, the spread of superradiant decay rates is considerably higher.\cite{pustovit-prl09}

  \begin{figure}
  \centering
  \includegraphics[width=0.8\columnwidth]{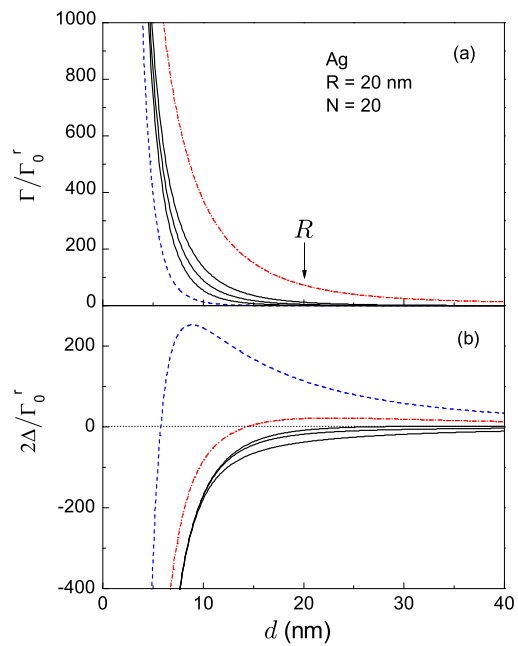}
  \caption{\label{fig:rate_20} (Color online) (a) Decay rates and (b) energy shifts vs. distance for 20 dipoles in C20 configuration around Ag NP. Each line corresponds to a system eigenstate and is similarly marked in both graphs. The dash-dotted line corresponds to three degenerate superradiant states, the dashed line is the darkest subradiant states dominated by dipole-dipole interactions, and solid lines correspond to the rest of subradiant states.}
  \end{figure}

\begin{figure}
  \centering
  \includegraphics[width=0.8\columnwidth]{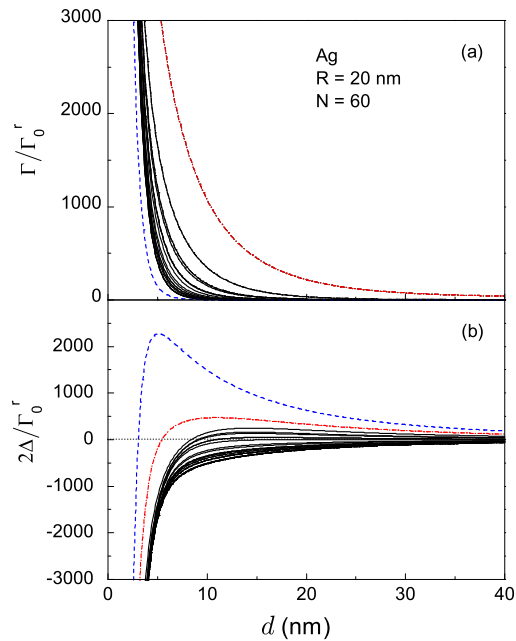}
  \caption{\label{fig:rate_60} (Color online) Same as in Fig.~\ref{fig:rate_20}, but for 60 dipoles in C60 configuration.}
  \end{figure}

\begin{figure}
  \centering
  \includegraphics[width=0.8\columnwidth]{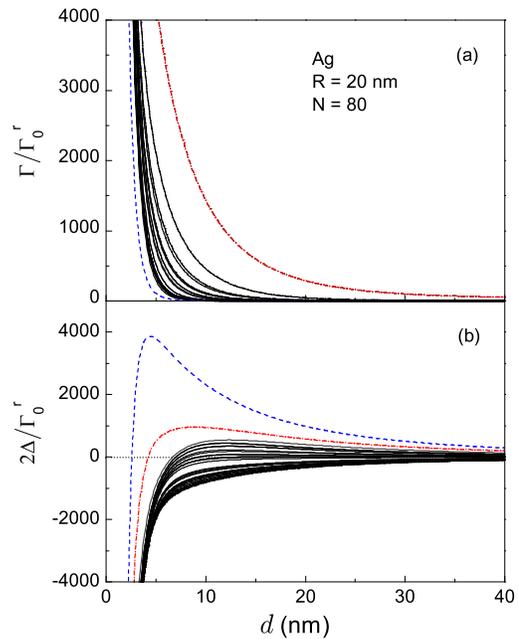}
  \caption{\label{fig:rate_80} (Color online) Same as in Fig.~\ref{fig:rate_20}, but for 80 dipoles in C80 configuration.}
  \end{figure}

To elucidate the structure of collective states, we calculate the distance dependences of complex eigenvalues,  $\Delta_{J}-\frac{i}{2}\Gamma_{J}$,  for C20, C60, and C80 configurations of dyes, as  shown in Figs.\ \ref{fig:rate_20}, \ref{fig:rate_60}, and \ref{fig:rate_80}, respectively.  For C20 configuration  (Fig.~\ref{fig:rate_20}), there are five sets of eigenvalues with 3, 4, 4, 7, and 1-fold degeneracies, in descending order of $\Gamma_{J}$ magnitudes. Down to the distance of $d=5$ nm, the largest decay rates, corresponding to three predominantly superradiant states, are well separated from the rest. The steep rise of $\Gamma_{J}$ at small distances is due to increasing contribution of high-$l$ plasmon modes close to NP surface [see Eqs.~(\ref{gamma-perp}) and (\ref{delta-perp})]. The interplay between various coupling mechanisms is especially revealing when comparing the plots for $\Delta_{J}$ and $\Gamma_{J}$ (curves for same eigenvalue sets have similar patterns). By their $d$ dependence, the eigenvalues fall into three main groups. The superradiant states have the largest decay rate $\Gamma_{J}$ for all $d$ and relatively small mainly \textit{positive} frequency shift for $d \gtrsim R/2$; these states are dominated by \textit{plasmon-enhanced radiative coupling}. The non-degenerate state with large \textit{positive} energy shift and smallest decay rate is dominated by direct nearest-neighbor dipoles interaction; this state is least affected by the presence of NP and does not participate in the emission. The third group of states with mostly \textit{negative}  $\Delta_{J}$ and small $\Gamma_{J}$ is dominated by \textit{nonradiative plasmon coupling}. Closer to NP surface, the coupling becomes dominant due to high-$l$ plasmons and all states develop large decay rates and negative energy shifts. Note that down to $d \gtrsim R/4$, the admixture between superradiant and subradiant modes is still relatively weak; below $R/4$, the non-radiative coupling dominates the spectrum and the admixture is strong. 
 

For larger ensembles, the eigenstates have similar structure, as illustrated in Figs.\ \ref{fig:rate_60} and \ref{fig:rate_80} which show calculated eigenvalues for dipoles in C60 and C80 configurations, respectively. Importantly, even with decreasing distance between the emitters in large ensembles, the dipole-dipole interactions still do not destroy cooperative emission. This can be understood from the following argument.\cite{shahbazyan-prb00} Mixing of superradiant and subradiant states takes place if the interactions between them are sufficiently strong. The latter requires that the electric field of a collective state is strongly inhomogeneous in space since, e.g., subradiant states couple only weakly to homogeneous field. On the other hand, such a field is comprised of individual fields of all the constituent dipoles so the resulting field's spatial fluctuations are weak if no two dipoles approach too close to each other, i.e., deviations of nearest-neighbor separations from their average, $\bar{s}=LN^{-1/3}$, $L$ being characteristic system size, are small. However, if deviations from $\bar{s}$ are large, i.e., two dipoles can be separated by a much closer distance, $s\ll LN^{-1/3}$, causing a strong spatial field fluctuation, then the eigenstates are no longer superradiant and subradiant states  and cooperative emission is destroyed. This argument was confirmed numerically here by finding system eigenstates for both cases -- dipoles on a spherical lattice with some fluctuations (see Figs.~\ref{fig:hist_delta-disorder} and  \ref{fig:hist_gamma-disorder}), and a completely random angular distribution with \textit{fixed} dipole-NP distance with no minimal separation between two dipoles (not shown). In the latter case, no superradiant states were formed and the reason was traced to configurations with extremely close dipoles. Note, however, that with both radial and angular distributions being random, these are rare events. In the case of repulsive interactions between individual dipoles, considered here, deviations from the average dipole-dipole separation $\bar{s}$ are exceedingly small and cooperative emission survives the interactions.

Another sharp contrast between plasmonic and photonic Dicke effects is the fate of subradiant states. In the latter, the energy trapped in subradiant states is eventually radiated, albeit with a much slower rate, resulting in sharp spectral features of emission spectrum.\cite{haroche-pr82,andreev-book} Instead, in plasmonic systems, the trapped energy is dissipated in the NP and only a small fraction of total energy leaves the system via superradoant states. Thus, the net effect of plasmonic Dicke effect is to drastically \textit{reduce} the emission as compared to same number of individual  dipoles. Remarkably, as the eigenvalues scale \textit{uniformly} with $N$, the \textit{quantum efficiencies} of superradiant states are nearly independent of the ensemble size, leading to the simple relation (\ref{energy}) that holds in the cooperative regime.

  \begin{figure}
  \centering
  \includegraphics[width=3.0in]{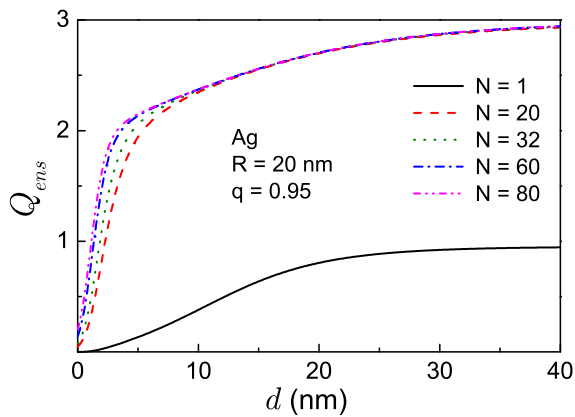}
  \caption{\label{fig:efficiency_q095} (Color online) Fluorescence quantum efficiency vs. distance for several ensembles of high-yield emitters on spherical lattices around AG NP.}
  \end{figure}

  \begin{figure}
  \centering
  \includegraphics[width=3.0in]{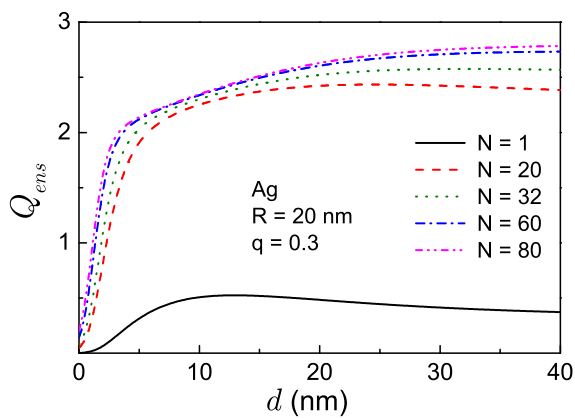}
  \caption{\label{fig:efficiency_q031} (Color online) Fluorescence quantum efficiency vs. distance for several ensembles of low-yield emitters on spherical lattices around AG NP.}
  \end{figure}

This is illustrated in  Figs.\ \ref{fig:efficiency_q095} and \ \ref{fig:efficiency_q031}, which show  \textit{ensemble} quantum efficiencies $Q_{ens}$ [see Eq.~(\ref{Q-ens})]  for two types of dyes with quantum yields $q=0.95$ and  $q=0.3$, respectively.  Two regimes can be clearly distinguished in the distance dependence of $Q_{ens}$: it first shows a sharp rise with its slope proportional to $N$ (non-cooperative regime) followed by a slower $d$ dependence (cooperative regime). The crossover between two regimes takes place at $d\simeq 5$ nm due to diminished high-$l$ plasmons contribution to nonradiative coupling for larger distances. In the cooperative regime, the precise behavior of $Q_{ens}$  is affected by molecules' quantum yield. For high-$q$ molecules, all $Q_{ens}$ dependences collapse onto a single curve, $Q_{ens}=3Q$, while for low-$q$ molecules, $Q_{ens}$ shows a weak dependence on $N$. In both cases, this behavior can be easily understood  from Eq.~(\ref{total_energy-ens}). Indeed, for degenerate superradiant eigenvalues we have $\lambda_{\mu}=1$, and for large distances, as $\Gamma \rightarrow \Gamma_{0}^{r}$, we obtain
\begin{align}
\label{Q-ens_far}
Q_{ens}\simeq \dfrac{3}{1+3(q^{-1}-1)/N},
\end{align}
i.e., for large ensembles, the role of molecular quantum yield is diminished.

\section{conclusions}
\label{sec:conc}

We studied here plasmon-mediated superradiance from an ensemble of dipoles near metal nanoparticle supporting localized surface plasmon, thereby extending the Dicke effect to plasmonic systems. Our main conclusion is that the plasmonic Dicke effect is a robust phenomenon, more so than the usual photonic Dicke effect, because of a more efficient hybridization of individual dipoles via nanoparticle plasmon.  We have established that hybridization takes place through two types of plasmonic coupling mechanisms -- plasmon-enhanced radiative coupling and nonradiative plasmon coupling, the latter having no analogue in the usual  Dicke effect and causing demise of cooperative emission at very close dipole-nanoparticle distances.

While we considered a specific nanostructure --  spherical metal particle -  the plasmon-mediated superradiance is a quite general phenomenon that should take place in any plasmonic system tuned into resonance with emitters, and Eq.~(\ref{energy}) should apply provided that the usual criteria for cooperative emission are met. In fact, our theory remains unchanged for any nanostructure with spherical symmetry, for example metal nanoshells with dielectric core, upon simple replacement of NP polarizabilities in self-energy matrix Eq.~(\ref{self}) and elsewhere with appropriate expressions. In fact, one expects that in nanoshells the nonradiatice losses would be smaller and so the plasmonic Dicke effect would be more robust than for solid nanoparticles.

This work was supported in part by the NSF under Grant Nos. DMR-0906945 and HRD-0833178, and under the EPSCOR Program.

\appendix
\section*{Appendix}
Here we collect relevant some formulas for the electric field Green dyadic in the presence of metal NP.  The Green dyadic satisfies Maxwell equation
\begin{eqnarray}
{\bm \nabla} \times {\bm\nabla} \times \hat{\bf G} - k^2 \epsilon(r) \hat{\bf G} = \hat{\bf I},
\end{eqnarray}
where $\epsilon(r) =\epsilon(\omega) \theta(R-r)+ \epsilon_{0}\theta(r-R)$ is local dielectric function ($\theta(x)$ is the step-function). The Green dyadic can be split into free space and Mie-scattered parts, $G_{\mu \nu }({\bf r},{\bf r}')=G^{0}_{\mu \nu }({\bf r},{\bf r}')+G^{s}_{\mu \nu }({\bf r},{\bf r'})$, where the free-space Green dyadic is
\begin{eqnarray}
G_{\mu\nu}^{0}({\bf r}-{\bf r'})
=\biggl( \delta_{\mu\nu} - \frac{ \nabla_{\mu} \nabla'_{\nu}}{k^2} \biggr) g({\bf r}-{\bf r}'),
\end{eqnarray}
with
\begin{eqnarray}
g({\bf r})=\frac{e^{ikr}}{4\pi r}
\end{eqnarray}
satisfying a scalar equation
\begin{eqnarray}
(\triangle  + k^2)g({\bf r})= -\delta({\bf r}).
\end{eqnarray}
Consider first the free space part. Its near field expression can be obtained in the long wave approximation, i.e. by expanding in $kr\ll 1$. In the first order, 
\begin{eqnarray}
\label{simplify}
G^{0}_{\mu\nu}({\bf r})=\frac{1}{4 \pi k^2 r^3} \Bigl[\frac{3 {\bf r}_{\mu} {\bf r}_{\nu}}{r ^2} - \delta_{\mu\nu} \Bigr]  + \dfrac{ik}{6\pi}\delta_{\mu\nu}.
\end{eqnarray} 
In the far field limit, i.e., $kr\gg1$ and $kr'\ll1$, the free-space part can be expanded via Bessel functions,
\begin{eqnarray}
\label{Bessel}
\frac{e^{ik|r-r'|}}{4\pi |r-r'|} = ik \sum_{lm} j_l (kr') h_l(kr) Y_{lm} (r) Y^{*}_{lm}(r'),
\end{eqnarray}
which are approximated as
\begin{eqnarray}
\label{approximation}
j_l(kr')=\frac{(kr')^l}{(2l+1)!!},
~~~
h_l(kr)=(-i)^{l+1} \frac{e^{ikr}}{kr},
\end{eqnarray}
yielding
\begin{align}
G_{\mu\nu}^{0}({\bf  r},{\bf  r}')=\Bigl(\delta_{\mu\nu}&-\frac{1}{k^2} {\nabla_{\mu} \nabla_{\nu}'}\Bigr)
\frac{e^{ikr}}{r} \biggl[\frac{1}{4\pi} 
\nonumber\\
&-  \frac{ikr'}{3} \sum_m Y_{1m} (\hat{\bf  r}) Y^{*}_{1m} (\hat{\bf  r}')\biggr].
\end{align}
After differentiation, the far field asymptotics takes the form
\begin{align}
G^{0}_{\mu\nu}({\bf  r},{\bf  r}')= \frac{e^{ik r}}{4\pi r} 
\biggl[ \delta_{\mu \nu } 
-\frac{4 \pi}{3} \sum_{m} \hat {\bf r}_{\mu}  Y_{1m}(\hat{\bf  r})\chi_{1m}^{\nu*}({\bf  r}')  \biggr],
\end{align}
where we introduced $\chi_{lm}^{\mu}({\bf r})=\nabla_{\mu} [r^{l} Y_{lm}(\hat{\bf r})$ and $\psi_{lm}^{\mu}({\bf r})=\nabla_{\mu} [r^{-l-1}Y_{lm}(\hat{\bf r})]$.

Now turn to the scattered part of the Green dyadic derived from solution of Mie problem for electromagnetic wave scattered on single sphere, 
\begin{eqnarray}
\label{green1}
G^{s}_{\mu \nu }({\bf r},{\bf r'},{\bf k})=i k \sum_{lm} \bigl[a_{l} N^{\mu}_{lm}({\bf r})N^{\nu }_{lm}({\bf r}') + 
\nonumber\\
b_{l} M^{\mu}_{lm}({\bf r})M^{\nu }_{lm}({\bf r}') 
\bigr],
\end{eqnarray}
where the first and second terms are electric and magnetic contributions and $a_{l}$ and  $b_{l}$ are the Mie coefficients. In the long wave approximation, $kR\ll 1$, the magnetic contribution in  Eq.(\ref{green1}) can be neglected as $b_{l}\ll 1$.\cite{nitzan-jcp81,ruppin-jcp82,ronis-jcp85,chew-jcp87} The Mie coefficient $a_{l}$ has a form
\begin{eqnarray}
a_{l}=\frac{\epsilon_{0} j_{l}(\rho_{0}) [\rho j_{l}(\rho)]'-
\epsilon j_{l}(\rho) [\rho_{0} j_{l}(\rho_{0})]'}
{\epsilon_{0} h_{l}(\rho_{0}) [\rho j_{l}(\rho)]'-
\epsilon j_{l}(\rho) [\rho_{0} h_{l}(\rho_{0})]'}
\end{eqnarray}
where $\rho_{i}=k_{i}R$, $k_i=\frac{\omega }{c} \sqrt{\epsilon_i}$, and $i=(\epsilon,\epsilon_0)$.  For $kR\ll 1$, it becomes
\begin{eqnarray}
\label{Mie coefficient}
a_{l}=-i s_{l} \tilde  \alpha_{l} k^{2l+1}, ~~ s_{l}=\frac{l+1}{l(2l+1) [(2l-1)!!]^2},
\end{eqnarray}
where 
\begin{eqnarray}
\label{alpha}
\tilde  \alpha_{l}=\frac {\alpha_{l}}{1-i s_{l} k^{2l+1} \alpha_{l}},
\end{eqnarray}
is NP multipolar polarizability that accounts for plasmon radiative decay,  and 
\begin{equation}
\alpha_{l}=R^{2l+1} \frac{l(\epsilon -\epsilon_0)}{l\epsilon +(l+1)\epsilon_0},
\end{equation}
is the standard NP polarizability. The function ${\bf N}_{lm}({\bf r})$ is given by 
\begin{eqnarray}
\label{N_function}
{\bf N}_{lm}({\bf r})
 =\frac{1}{k\sqrt{l(l+1)}}
{\bm \nabla} \times \bigl [h^{(1)}_{l}(kr)  {\bf L} Y_{lm}(\hat{\bf r})\bigl ],
\end{eqnarray}
where $ {\bf L} = -i ({\bf r} \times {\bm \nabla})$ is angular momentum operator. Using the following identity,
\begin{align}
\label{property}
{\bm \nabla} \times \bigl [h^{(1)}_{l} & (kr) {\bf L} Y_{lm}(\hat{\bf r})\bigl ]
 = i {\bf r} k^2 h^{(1)}_l(kr) Y_{lm}(\hat{\bf r})
 \nonumber\\
 & + i  {\bm \nabla} \bigl[[kr h^{(1)\prime}_{l}(kr) + h^{(1)}_{l}(kr)] Y_{lm}(\hat{\bf r})\bigl],  
\end{align}
prime standing for derivative, and expanding $h^{(1)}_{l}(kr)=j_{l}(kr)+i n_{l}(kr)$ in $kr$ as 
\begin{eqnarray}
\label{h-expansion}
h^{(1)}_{l}(kr)= \frac{(kr)^{l}}{(2l+1)!!} - i \frac{(2l-1)!!}{(kr)^{l+1}}, 
\end{eqnarray}
we obtain
\begin{eqnarray}
\label{N-close}
{\bf N}_{lm}(kr)= - \frac{1}{k \sqrt{s_{l} (2l+1)}} {\bm \nabla} [\varphi _{l}(kr) Y_{lm}(\hat{\bf r})],
\end{eqnarray}
where 
\begin{eqnarray}
\varphi _{l}(kr) = \frac{1}{(kr)^{l+1}} - i s_{l} (kr)^{l}.
\end{eqnarray}
Thus, for $kr\ll 1$ and $kr'\ll 1$, the scattered part of the Green dyadic has the form
\begin{align}
&G^{s}_{\mu \nu }({\bf r},{\bf r'},{\bf k})
\approx
i k  \sum_{lm} \bigl[a_{l} N^{\mu}_{lm}(kr)N^{\nu }_{lm}(kr')\bigr]
\\
&\approx \sum_{lm} 
\frac {k^{2l} \tilde \alpha_{l}}{2l+1} \nabla_{\mu } \bigl[\varphi _{l}(kr) Y_{lm}(\hat{\bf r})\bigr]
\nabla'_{\nu} \bigl[\varphi _{l}(kr')Y^{*}_{lm}(\hat{\bf r}')\bigr].
\nonumber
\end{align}
This expression can be further simplified by substituting $\tilde \alpha_{l}=\bar \alpha_{l}+ i s_{l} k^{2l+1} |\tilde\alpha_{l}|^2$, where $\bar \alpha_{l}= \alpha_{l}|1-is_{l} k^{2l+1}\alpha_{l}|^{-2}$, and  keeping the first two powers of $k$
\begin{align}
G^{s}_{\mu \nu }({\bf r},{\bf r'})
=\dfrac{1}{k^2}\sum_{lm} \frac{\bar  \alpha_{l}}{(2l+1)} 
\psi_{lm}^{\mu}({\bf r})\psi_{lm}^{\nu *}({\bf r}') &
\nonumber\\
-  i\frac{ks_{1}}{3} \sum_{m=-1}^{1} \Bigl[\tilde \alpha_{1} \bigl[\psi_{1m}^{\mu}({\bf r})\chi_{1m}^{\nu *}({\bf r}')
+  \chi_{1m}^{\mu} &({\bf r})\psi_{1m}^{\nu *}({\bf r}')\bigr] 
\nonumber\\
-|\tilde \alpha_{1}|^2 \psi_{1m}^{\mu}({\bf r})\psi_{1m}^{\nu*}({\bf r}')\Bigr],&
\end{align}
which, after adding the free-space part of the Green dyadic and neglecting plasmon radiative decay, leads to Eq.~(\ref{self}).


For $kr\gg1$, with help of Eqs. (\ref{approximation}), (\ref{N_function}), and (\ref{property}), we easily obtain
\begin{eqnarray}
\label{N-function_far}
{\bf N}_{lm}({\bf r})=-\frac{(-i)^{l+1} e^{ikr}}{k \sqrt{l(l+1)}} {\bm \nabla} Y_{lm}(\hat{\bf r}),
\end{eqnarray}
and combining this expression with  Eq.~(\ref{N-close}), we obtain the far field Green dyadic (i.e., $kr\gg 1$ and $kr'\ll 1$)
\begin{align}
\label{Green_far}
G^{s}_{\mu \nu }({\bf r},{\bf r'})=-\frac{\tilde \alpha_1}{3} \sum_{m=-1}^{1}   e^{ikr} \bigl[ \nabla_{\mu} Y_{1m}(\hat{\bf r})\bigr]  \psi_{1m}^{\nu*}({\bf r'}),
\end{align}
where we set $l=1$.

{}

\end{document}